\documentclass[conference,letterpaper]{IEEEtran}

\pdfoutput=1

\ifCLASSINFOpdf
   \usepackage[pdftex]{graphicx}
\else
  \usepackage[dvips]{graphicx}
\fi
\usepackage[cmex10]{amsmath}

\def\(({\left(}
\def\)){\right)}
\def\[[{\left[}
\def\]]{\right]}
\newcommand{\be}{\begin{equation}}
\newcommand{\ee}{\end{equation}}
\newcommand{\bea}{\begin{eqnarray}}
\newcommand{\eea}{\end{eqnarray}}

\newcommand{\bF}{{\textbf {F}}}

\newcommand{\bx}{{\textbf {x}}}

\newcommand{\bs}{{\textbf {s}}}
\newcommand{\by}{{\textbf {y}}}

\begin{document}
\title{Compressed Sensing of Approximately-Sparse Signals: Phase
  Transitions and Optimal Reconstruction}

\author{\IEEEauthorblockN{Jean Barbier\IEEEauthorrefmark{1}, Florent Krzakala\IEEEauthorrefmark{1},
Marc M\'ezard\IEEEauthorrefmark{2}, and
Lenka Zdeborov\'a\IEEEauthorrefmark{3}}
\IEEEauthorblockA{
\IEEEauthorrefmark{1}
CNRS and  ESPCI ParisTech 10 rue Vauquelin, UMR   7083 Gulliver 
Paris 75005  France}
\IEEEauthorblockA{
\IEEEauthorrefmark{2}
Univ. Paris-Sud \& CNRS, LPTMS, UMR8626  B\^{a}t.~100, 91405 Orsay, France.}
\IEEEauthorblockA{
\IEEEauthorrefmark{3}
Institut de Physique Th\'eorique, IPhT, CEA Saclay, and URA 2306
CNRS, 91191 Gif-sur-Yvette, France}
}

% make the title area
\maketitle

\begin{abstract}
  Compressed sensing is designed to measure sparse signals directly in
  a compressed form. However, most signals of interest are only
  ``approximately sparse'', i.e. even though the signal contains only
  a small fraction of relevant (large) components the other components
  are not strictly equal to zero, but are only close to zero. In this
  paper we model the approximately sparse signal with a Gaussian
  distribution of small components, and we study its compressed sensing with dense random
  matrices. We use replica calculations to determine the mean-squared
  error of the Bayes-optimal reconstruction for such signals, as a
  function of the variance of the small components, the density of large
  components and the measurement rate. We then use the G-AMP algorithm 
and we quantify the region of parameters for which this
  algorithm achieves optimality (for large systems). Finally, we show
  that in the region where the G-AMP for the homogeneous measurement
  matrices is not optimal, a special ``seeding'' design of a
  spatially-coupled measurement matrix allows to restore optimality.
\end{abstract}

\section{Introduction}

Compressed sensing is designed to measure sparse signals directly in a
compressed form. It does so by acquiring a small number of random
linear projections of the signal and subsequently reconstructing the
signal. The interest in compressed sensing was boosted by
works \cite{CandesTao:06,Donoho:06} that showed that this reconstruction is computationally
feasible in many cases. Often in the studies of compressed sensing the
authors require that the reconstruction method works with guarantees
for an arbitrary signal. This requirement then has to be paid by
higher measurement rates than would be necessary if the probabilistic
properties of the signal were at least approximately known. In many
situations where compressed sensing is of practical interest there is
a good knowledge about the statistical properties of the signal. In the present paper we
will treat this second case. 

It has been shown recently \cite{KrzakalaPRX2012,DonohoJavanmard11} that for compressed sensing of sparse
signals with known empirical distribution of components the theoretically optimal
reconstruction can be achieved with the combined use of G-AMP algorithm \cite{DonohoMaleki10,Rangan10b}
and seeding (spatially coupled) measurement matrices. Actually, \cite{KrzakalaPRX2012}
argued that for noiseless measurements the knowledge of the signal
distribution is not even required. It is well known that for noiseless measurements exact
reconstruction is possible down to measurement rates equal to the density of non-zero components in the signal. For noisy
measurements the optimal achievable mean-squared error (MSE) has been analyzed and compared to
the performance of G-AMP in \cite{KrzakalaMezard12}.  

In its most basic form compressed sensing is designed for sparse
signals, but most signals of interest are only ``approximately
sparse'', i.e. even though the signal contains only a small fraction
of relevant (large) components the other components are not strictly
equal to zero, but are only close to zero. In this paper we model the
approximately sparse signal by the two-Gaussian distribution as in
\cite{BaronSarvotham10}. We study the optimal achievable MSE in the
reconstruction of such signals and compare it to the performance of
G-AMP algorithm using its asymptotic analysis - the state evolution
\cite{DonohoMaleki10,Rangan10b,KrzakalaMezard12}.  Even-though we
limit ourselves to this special class of signals and assume their
knowledge, many qualitative features of our results stay true for
other signals and when the distribution of signal-components is
not known as well.

\subsection{Definitions}

We study compressed sensing for approximately sparse signals. The
$N$-dimensional signals that we consider have iid components, $K$ of
these components are chosen from a distribution $\phi(x)$, we define
the density $\rho=K/N$, and the remaining $N-K$ components are
Gaussian with zero mean and small variance $\epsilon$ \be P(\bx) =
\prod_{i=1}^N [ \rho\phi(x_i) + (1-\rho) {\cal N}(0, \epsilon)
] \label{Px} \ee Of course no real signal of interest is iid, however,
for the same reasons as in \cite{DonohoJavanmard11} our analysis
applies also to non-iid signals with empirical distribution of
components converging to $\rho\phi(x) + (1-\rho) {\cal N}(0, \epsilon)
$. For concreteness our numerical examples are for Gaussian $\phi(x)$
of zero mean and unit variance. Although the numerical results depend
on the form of $\phi(x)$, the overall picture is robust with respect
to this choice. We further assume that the parameters of $P(\bx)$ are
known or can be learned via expectation-maximization-type of approach.

We obtain $M$ measurements $y_\mu$ as linear projections of a
$N$-components signal
\be
    y_\mu  = \sum_{i=1}^N F_{\mu i }s_i\, , \quad \quad
    \mu=1,\dots,M\, .\label{def}
\ee
The $M\times N$ measurement matrix is denoted $F_{\mu i}$. For
simplicity we assume the measurements to be noiseless, the case of noisy measurements can be treated in the same way as in
\cite{KrzakalaMezard12}.  As done traditionally, in the first part of
this paper, we consider the measurement matrix having iid components of zero mean
and variance $1/N$. In our numerical experiments we chose the
components of the matrix to be normally distributed, but the
asymptotic analysis does not depend on the details of the components
distribution. The seeding measurement matrices considered in the second part of this paper
will be defined in detail in section~\ref{seeding}.  

The goal of compressed sensing is to reconstruct the signal $\bs$
based on the knowledge of $M$ measurements $\by$ and the $M\times N$
matrix $\bF$. We define $\alpha=M/N$ to be the measurement (sampling)
rate.  The Bayes optimal way of estimating a signal $\bx^{\star}$ that
minimizes the MSE $E=\sum_{i=1}^N (s_i - x_i^{\star})^2 /N$ with the
true signal $\bs$ is given as \be x^{\star}_i=\int \text{d} x_i \,
x_i\, \nu_i(x_i) \, ,\label{average_marginal} \ee where $\nu_i(x_i)$
is the marginal probability distribution of the variable $i$ 
\be
\nu_i(x_i) \equiv \int_{\{x_j\}_{j\neq i}} P(\bx| \by)  \prod_{j \neq
  i} dx_j\ee
 under the
posterior measure \be P(\bx | \by) = \frac{1}{Z(\by)} P(\bx)
\prod_{\mu=1}^M \delta(y_\mu - \sum_{i=1}^N F_{\mu i}x_i)\,
.\label{p_bayes} \ee In this paper we will use an asymptotic replica
analysis of this optimal Bayes reconstruction, which allows to compute
the MSE as function of the parameters of the signal distribution,
$\rho$ and $\epsilon$, and of the measurement rate $\alpha$.

Of
course optimal Bayes reconstruction is not computationally
tractable. In order to get an estimate of the marginals $\nu_i(x_i)$,
we use the G-AMP algorithm that is a belief-propagation based
algorithm. 

\subsection{Related works}
The $\ell_1$-minimization based algorithms \cite{CandesTao:06,Donoho:06} are widely used
for compressed sensing of approximately sparse signals. They are very
general and provide good performances in many situations. They,
however, do not achieve optimal reconstruction when the statistical
properties of the signal are known.  

The two-Gaussian model for approximately sparse signal eq.~(\ref{Px})
was used in compressed sensing e.g. in \cite{BaronSarvotham10,KudekarPfister10}. 

Belief propagation based reconstruction algorithms were introduced in
compressed sensing by \cite{BaronSarvotham10}. Authors of \cite{BaronSarvotham10} used sparse
measurement matrices and treated the BP messages as probabilities over
real numbers, that were represented by a histogram. The messages,
however, can be represented only by their mean and variance as done by
\cite{GuoWang06,Rangan10}. Moreover, one does not need to send messages
between every signal-components and every measurements
\cite{DonohoMaleki09}, this leads to the approximate message passing
(AMP). 
In the context of physics of spin glasses this transformation of the
belief propagation equations corresponds to the
Thouless-Anderson-Palmer equations \cite{ThoulessAnderson77}.
The AMP was generalized for general signal models in
\cite{DonohoMaleki10,Rangan10b} and called G-AMP. The algorithm used
in \cite{KrzakalaPRX2012,KrzakalaMezard12} is equivalent to G-AMP. We
also want to note that we find the name ``approximate'' message
passing a little misleading since, as argued e.g. in \cite{KrzakalaMezard12},
for dense random measurement matrices the G-AMP is asymptotically
equivalent to BP, i.e. all the leading terms in $N$ are included in
G-AMP.

For random matrices the evolution of iterations of G-AMP on large
system sizes is described by state evolution
\cite{DonohoMaleki09}. The exactness of this description was proven in
large generality in \cite{BayatiMontanari10}. See also
\cite{GuoWang07,Rangan10b,KrzakalaMezard12} for discussions and results
on the state evolution.

The optimal reconstruction was studied extensively in
\cite{WuVerdu11}. The replica method was used to analyse the optimal
reconstruction in compressed sensing in e.g. \cite{RanganFletcherGoyal09,GuoBaron09}.
In the statistical physics point of view the
replica method is closely related to the state evolution \cite{KrzakalaMezard12}.

As we shall see the G-AMP algorithm for homogeneous measurement
matrices matches asymptotically the performance of the optimal
reconstruction in a large part of the parameter space. In some region
of parameters, however, it is suboptimal.  For the sparse signals, it
was demonstrated heuristically in \cite{KrzakalaPRX2012} that
optimality can be restored using seeding matrices (the concept is
called spatial coupling), rigorous proof of this was worked out in
\cite{DonohoJavanmard11}. The robustness to measurement noise was also
discussed in \cite{DonohoJavanmard11,KrzakalaMezard12}. Note that the
concept of ``spatial coupling'' thanks to which theoretical thresholds
can be saturated was developed in error-correcting codes
\cite{FelstromZigangirov99,KudekarRichardson10,KudekarRichardson12}.
In compressed sensing the ``spatial coupling'' was first tested in
\cite{KudekarPfister10} who did not observe any improvement for the
two-Gaussian model for reasons that we will clarify later in this
paper. Basically, the spatial coupling provides improvements only if a
first order phase transition is present, but for the variance of small
components that was tested in \cite{KudekarPfister10} there is no such
transition: it appears only for slightly smaller values of the
variance.

\subsection{Our Contribution}
Using the replica method, we study the MSE in optimal
Bayes inference of approximately sparse signals. In parallel, we study
the asymptotic (large $N$)  performance of G-AMP using state evolution. The parameters that we vary
are the density of large components $\rho=K/N$, the variance of the small
components $\epsilon$ and the sampling rate $\alpha=M/N$.

More precisely, we show that for a fixed signal density
$\rho$, for low variance of the small components
$\epsilon<\epsilon(\rho)$, the optimal Bayes reconstruction has a
 transition at a critical value $\alpha=\alpha_{\rm opt}$,
separating a phase with a small value (comparable to $\epsilon$) of
the MSE, obtained at  $\alpha>\alpha_{\rm opt}$, from a phase with a large value of
the MSE, obtained at  $\alpha<\alpha_{\rm opt}$. This is a ``first
order'' phase transition, in the sense that the MSE is discontinuous
at $\alpha=\alpha_{\rm opt}$.

The G-AMP algorithm exhibits a double phase transition. It
 is asymptotically equivalent to the optimal Bayes inference at large $\alpha_{\rm BP}<\alpha<1$,
where it matches the optimal reconstruction with a small value of the MSE. At
low values of $\alpha<\alpha_{\rm opt}$ the G-AMP is also
asymptotically equivalent to the optimal Bayes inference but in this
low-sampling-rate region the optimal result  leads to a
large MSE. In the intermediate region $\alpha_{\rm opt}<
\alpha<\alpha_{\rm BP}$ G-AMP leads to large MSE, but the optimal
Bayes inference leads to low MSE. This is the region where one needs to improve on G-AMP.
We show that in this intermediate
region the G-AMP performance can be improved with the use of seeding
(spatially coupled) measurement matrices, and with a proper choice of the parameters
of these matrices one can approach the performance of the optimal
Bayes inference in the large system size limit. 

Finally for higher
variance of the small components $\epsilon>\epsilon(\rho)$ there is no
phase transition for $0<\alpha<1$. In this regime, G-AMP achieves optimal
Bayes inference and the MSE that it obtains varies continuously from 0 at $\alpha=1$ to
large values at low measurement rate $\alpha$.

\section{Bayes optimal and G-AMP reconstruction of approximately sparse signals}

If the only available information about the signal is the matrix $\bF$
and the vector of measurements $\by$ then the
information-theoretically best possible estimate of each signal
component is computed as a weighted average over all solutions of the
linear system (\ref{def}), where the weight of each solution is given
by~(\ref{Px}). Of course, the undetermined linear system (\ref{def})
has exponentially many (in $N$) solutions and hence computing exactly
the above weighted average is in general intractable.

The corresponding expectation can be, however, approximated
efficiently via the generalized approximate message passing (G-AMP)
algorithm
\cite{Rangan10b,DonohoMaleki10,KrzakalaPRX2012,KrzakalaMezard12} that
we recall in Sec.~\ref{GAMP}. The behavior of the algorithm in the
limit of large system sizes can be analyzed via state
evolution \cite{BayatiMontanari10,KrzakalaPRX2012,KrzakalaMezard12}, as we recall in Sec.~\ref{evolution}.

The asymptotic performance of the optimal reconstruction can be analyzed
via the replica method as in \cite{RanganFletcherGoyal09,GuoBaron09,KrzakalaMezard12},
which is closely related to the state evolution of the G-AMP
algorithm. We summarize the corresponding equations in Sec.~\ref{replicas}.

\subsection{Reminder of the G-AMP Algorithm}
\label{GAMP}

The G-AMP is an iterative message passing algorithm. For every
measurement component we define quantities $V_\mu$ and $\omega_\mu$,
for each signal component quantities $\Sigma_i$, $R_i$, $a_i$,
$v_i$. In the G-AMP algorithm these quantities are updated as follows
(for the derivation in the present notation see
\cite{KrzakalaMezard12}, for the original derivation \cite{Rangan10b,DonohoMaleki10})
\bea
 V^{t+1}_\mu &=& \sum_i F_{\mu i}^{2} \, v^{t}_i \, , \label{TAP_ga} \\
\omega^{t+1}_\mu  &=& \sum_i F_{\mu i} \, a^t_i -\frac{
  (y_\mu-\omega^t_\mu)}{V^t_\mu} \sum_i F_{\mu i}^2\,
v^t_i \, , \label{TAP_al}  \\
   (\Sigma^{t+1}_i)^2&=&\left[ \sum_\mu \frac{F^2_{\mu i}}{ V^{t+1}_\mu} \right]^{-1}\, ,
      \label{TAP_U}\\
      R^{t+1}_i&=& a^t_i + \frac{\sum_\mu F_{\mu i} \frac{(y_\mu - \omega^{t+1}_\mu)}{ V^{t+1}_\mu}}{ \sum_\mu \frac{ F_{\mu i}^2}{V^{t+1}_\mu}}\, , \label{TAP_V}\\
      a^{t+1}_i &=&   f_a\left((\Sigma^{t+1}_i)^2,R^{t+1}_i\right) ,  \label{TAP_a}\\
      v^{t+1}_i &=& f_c\left((\Sigma^{t+1}_i)^2,R^{t+1}_i\right) \, . \label{TAP_v}
\eea
Here only the functions $f_a$ and $f_b$ depend in an explicit way on
the signal model $P(\bx)$. For the signal model (\ref{Px}) considered
in this paper we have 
\begin{align}
&f_a(\Sigma^2,R) = \frac{  \sum_{a=1}^2 w_a
e^{-\frac{R^2}{2(\Sigma^2+\sigma_a^2)}}
\frac{ R \sigma_a^2}{(\Sigma^2+\sigma_a^2)^{\frac{3}{2}}} }{   \sum_{a=1}^2 w_a \frac{1}{\sqrt{\Sigma^2+\sigma_a^2}}
e^{-\frac{R^2}{2(\Sigma^2+\sigma_a^2)}} } \,
,\\
&f_b(\Sigma^2,R) = \frac{  \sum_{a=1}^2 w_a
e^{-\frac{R^2}{2(\Sigma^2+\sigma_a^2)}}
\frac{\sigma_a^2
  \Sigma^2 (\Sigma^2 + \sigma_a^2)+  R^2 \sigma_a^4 }{(\Sigma^2+\sigma_a^2)^{\frac{5}{2}}} }{    \sum_{a=1}^2 w_a \frac{1}{\sqrt{\Sigma^2+\sigma_a^2}}
e^{-\frac{R^2}{2(\Sigma^2+\sigma_a^2)}} } \nonumber \,
, \\
&f_c(\Sigma^2,R)=f_b(\Sigma^2,R)-f_a^2(\Sigma^2,R) \, .
\end{align}
For the approximately sparse signal that we consider in this paper we have
\bea
w_1=\rho\, , \quad
\sigma_1^2=\sigma^2\, , \\
w_2=1-\rho\, ,\quad \sigma_2^2=\epsilon\, .  
\eea
A suitable initialization for the quantities is $a_i^{t=0}=0$,
$v_i^{t=0}=(1-\rho)\epsilon + \rho \sigma^2 $,
$\omega_\mu^{t=0}=y_\mu$.

Once the convergence of the iterative equations is reached, i.e. the
quantities do not change anymore under iterations, the estimate of the
$i$th signal component is $a^{t}_i$.  The MSE achieved by the
algorithm is then $E^t=\sum_{i=1}^N (a_i^t - s_i)^2/N$.

\subsection{Evolution of the algorithm}
\label{evolution}

\begin{figure}[!ht]
\includegraphics[width=3.5in]{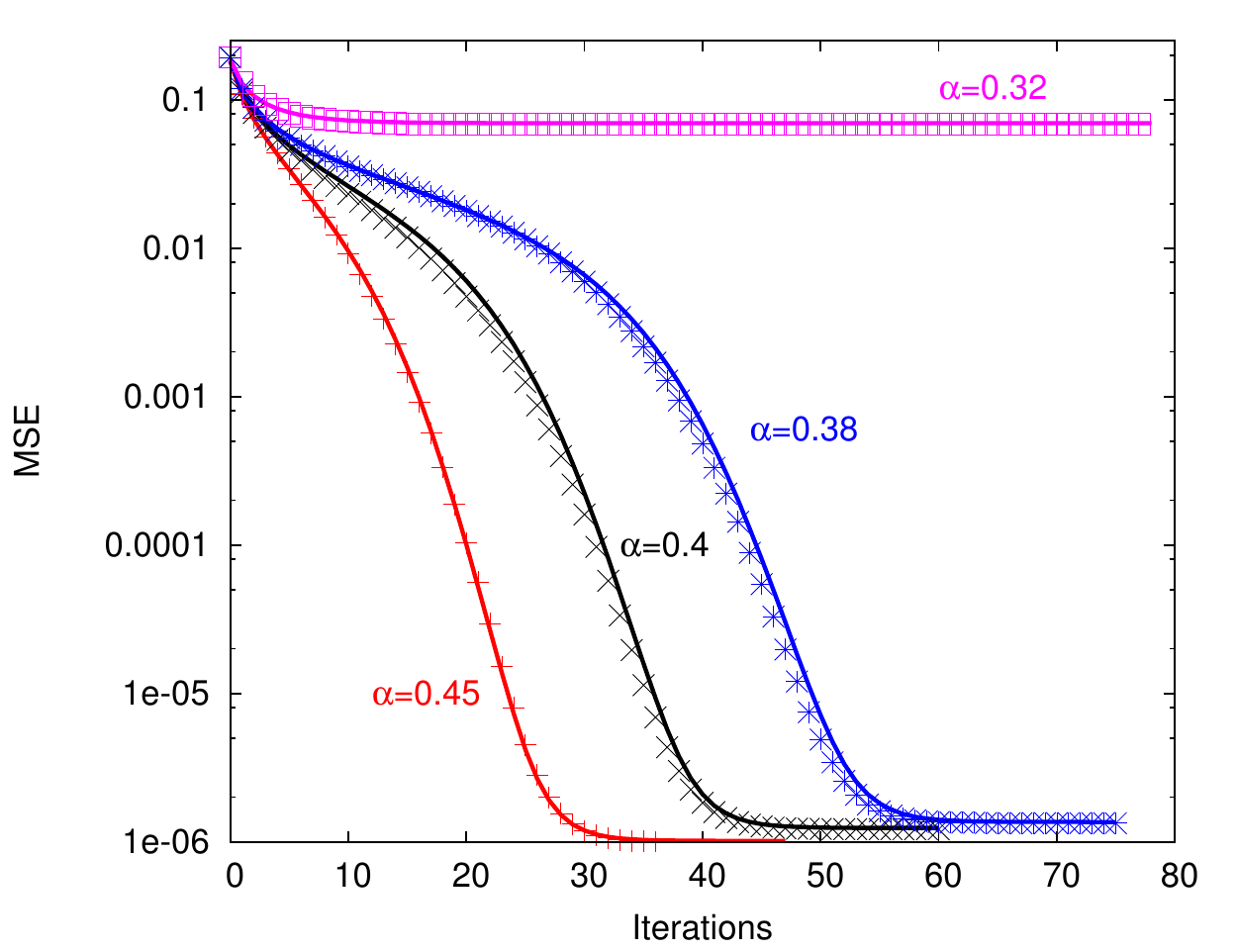}
\caption{Time-evolution of the MSE the G-AMP algorithm achieves (points)
  compared to the asymptotic $N\to \infty$ evolution obtained from the state
  evolution eq.~(\ref{Et}) (full lines). Data are obtained for a signal with
  density of large component $\rho=0.2$, variance of the small
  components $\epsilon=10^{-6}$. The algorithm was used for a signal
  of $N=3\cdot 10^4$ components.}
\label{fig_evolution}
\end{figure}

In the limit of large system sizes, i.e. when parameters $\rho,
\epsilon,\alpha$ are fixed whereas $N\to \infty$, the evolution of the
G-AMP algorithm can be described exactly using the  ``state
evolution'' \cite{BayatiMontanari10}. In the case where the signal
model corresponds to the statistical properties of the actual signal,
as it is the case in the present work, the state evolution is stated in
terms of a single variable $E^t$, the MSE at iteration-time $t$, which
evolves as (for a derivation see e.g. \cite{BayatiMontanari10,Rangan10b,KrzakalaMezard12})
\begin{align}
    &E^{t+1} = \sum_{a=1}^{2} w_a  \int {\cal D}z f_c\left( \frac{1}{\hat
      m^t}, z\sqrt{\sigma_a^2  + \frac{1}{\hat m^t} }\right)\, , \label{Et}\\
   &{\hat m}^t = \frac{\alpha}{E^{t}}\, ,
\end{align}
where ${\cal D} z = e^{-z^2/2}/\sqrt{2\pi} dz$ is a Gaussian measure
for the integral.  The initialization corresponding to the one for the
algorithm is $E^{t=0}=(1-\rho)\epsilon + \rho \sigma^2$.

In Fig.~\ref{fig_evolution} we plot the analytical prediction for the
time evolution of the MSE computed from the state evolution
(\ref{Et}),  and we compare it to the one measured in one run of the
G-AMP algorithm for a system size $N=3\cdot 10^4$. The agreement for
such system size is already excellent.

\subsection{Optimal reconstruction limit}
\label{replicas}

We notice that for some measurement rates $\alpha$ the state evolution
equation (\ref{Et}) has two different stable fixed points. In
particular, if the iterations are initialized with $E\to 0$, for certain
values of $\alpha$ one will reach a fixed point with much lower MSE
than initializing with large $E$. In fact, one of the fixed points
determines the MSE that would be achieved by the exact Bayes optimal
inference. This can be seen using the heuristic replica method that
leads to asymptotically exact evaluation of the logarithm of the
partition function $Z$ in eq.~(\ref{p_bayes}). In general, if the
partition function can be evaluated precisely then the expectations
$x_i^{\star}$ eq.~(\ref{average_marginal}) and the associated MSE of
the optimal inference can be computed.

\begin{figure}[!ht]
\includegraphics[width=3.5in]{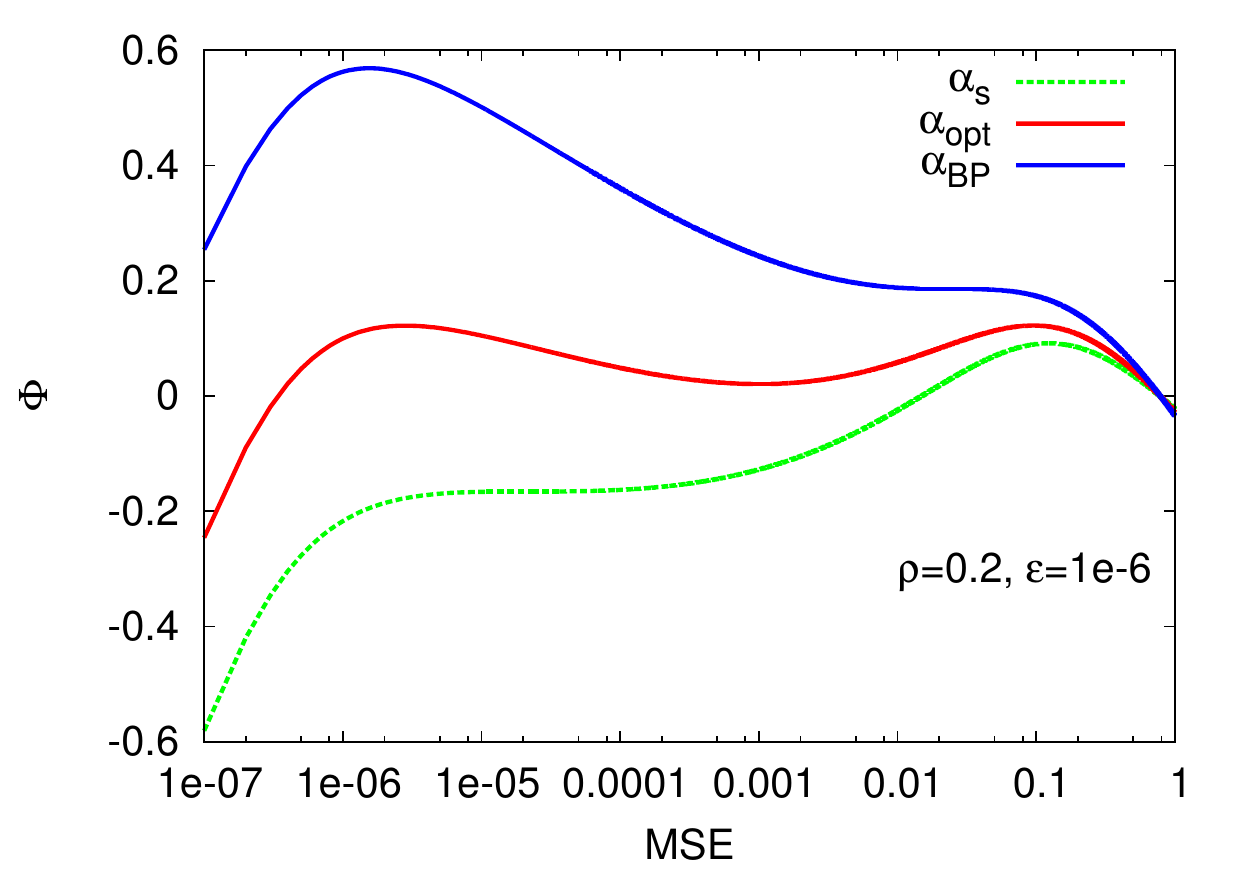}
\caption{The potential function $\Phi(E)$ for signals of density $\rho=0.2$, with
  variance of the small components $\epsilon= 10^{-6}$. The
  three lines depict the potential for three different measurement rates corresponding to the critical values: 
  $\alpha_{\rm BP}=0.3559$, $\alpha_{\rm opt}=0.2817$,
  $\alpha_s=0.2305$. The two local maxima exists for $\alpha \in
  (\alpha_s,\alpha_{\rm BP})$, at $\alpha_{\rm opt}$ the low MSE
  maxima becomes the global one.}
\label{fig_potential}
\end{figure}

The replica analysis, derived for the present problem e.g. in
\cite{KrzakalaMezard12}, shows  that the large $N$ limit of $\log{Z}/N$ is equal to the global maximum of the following
``potential'' function 
\begin{align}
\Phi(E) = &- \frac{\alpha}{2} \left(\log E + \frac{w_1\sigma_1^2 + w_2\sigma_2^2}{E}\right)
\nonumber \\ & +
\sum_{a=1}^2 w_a\int {\cal D}z \log{ \left[ \sum_{b=1}^2 w_b
    \frac{e^{\frac{ (\hat m^2 \sigma_a^2 + \hat m)z^2 }{ 2(\hat m+1/\sigma_b^2) } }}{\sqrt{\hat m \sigma_b^2 +1}}  \right]}
\end{align}
Note that the state evolution corresponds to the steepest ascent of
$\Phi(E)$. When $\Phi(E)$ has two local maxima then the fixed point of
(\ref{Et}) depends on the initial condition.

In Fig.~\ref{fig_potential} we plot the function $\Phi(E)$ for a signal
of density $\rho=0.2$, variance of small components $\epsilon=10^{-6}$
and three different values of the measurement rate $\alpha$. We
define three phase transitions 
\begin{itemize}
  \item{$\alpha_{\rm BP}$ is defined as the largest $\alpha$ for which the
      potential function $\Phi(E)$
      has two local maxima.}
\item{$\alpha_{s}$ is defined as the smallest $\alpha$ for which the potential
    function $\Phi(E)$ has two local maxima.}
  \item{$\alpha_{\rm opt}$ is defined as the value of $\alpha$ for which the two
      maxima have the same height.}
\end{itemize}

\section{Phase diagrams for approximate sparsity}
\label{results}

\begin{figure}[!ht]
\includegraphics[width=3.5in]{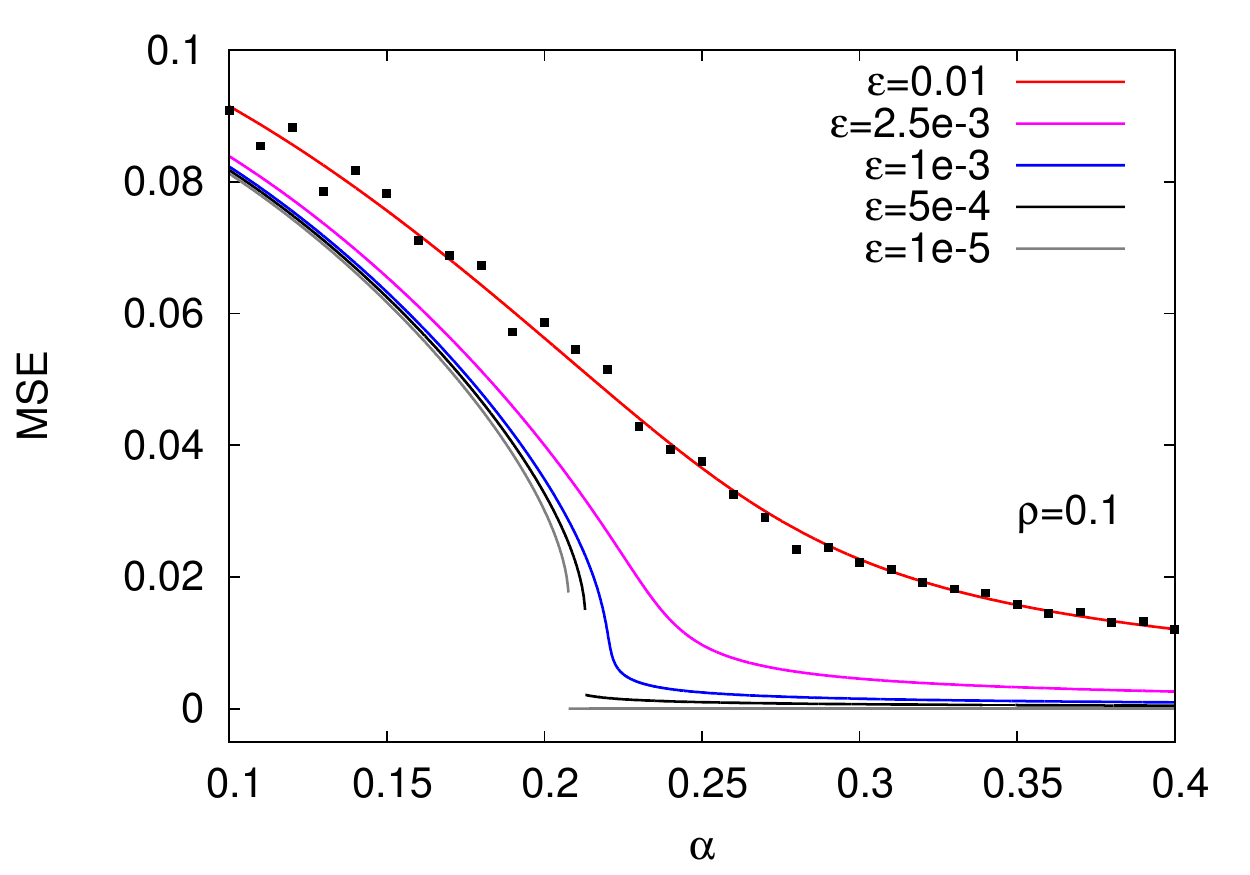}
\caption{The MSE achieved by the G-AMP. The lines correspond to the
  evaluation of the MSE from the state evolution, the data points to
  the MSE achieved by the G-AMP algorithm for $N=3\cdot 10^4$. The
  data are for signals with density $\rho=0.1$ and several values of
  variance of small components $\epsilon$ as a function of the
  measurement rate $\alpha$. The MSE grows continuously as $\alpha$
  decreases for $\epsilon>0.00075$. For smaller values of the noise a
  first order phase transition is present and the MSE jumps
  discontinuously at $\alpha_{\rm BP}(\epsilon)$. }
\label{fig_MSE1}
\end{figure}

In Fig.~\ref{fig_MSE1} we plot the MSE to which the state evolution
converges if initialized at large value of MSE - such initialization
corresponds to the iterations of G-AMP when the actual signal is not
known. For $\epsilon=0.01$ we also compare explicitly to a run of
G-AMP for system size of $N=3\cdot 10^4$. Depending on the value of density $\rho$, and variance
$\epsilon$, two situations are possible: For relatively large
$\epsilon$, as the measurement rate $\alpha$ decreases the final MSE
grows continuously from $E=0$ at $\alpha=1$ to $E=E^{t=0}$ at
$\alpha=0$. For lower values of $\epsilon$ the MSE achieved by G-AMP
has a discontinuity at $\alpha_{\rm BP}$ at which the second maxima of
$\Phi(E)$ appears. Note that the case of $\epsilon=0.01$ was tested in
\cite{BaronSarvotham10}, the case of $\epsilon=0.0025$ in \cite{KudekarPfister10}.

\begin{figure}[!ht]
\includegraphics[width=3.5in]{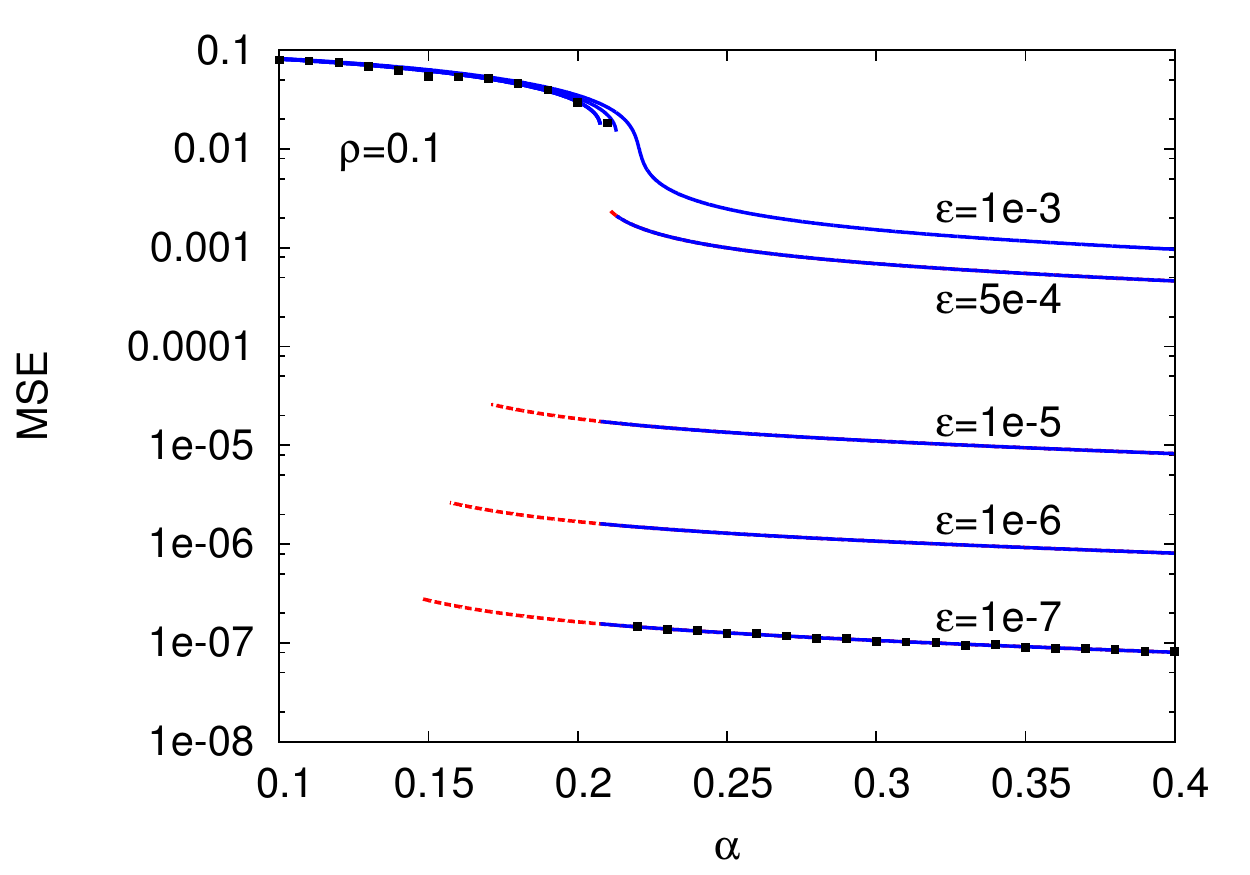}
\caption{MSE achieved by the G-AMP (blue solid lines) compared to the MSE
  achieved by the Bayes-optimal inference (red dashed lines) as
  evaluated using the state evolution. The data points correspond to
  the MSE achieved by the G-AMP algorithm for $N=3\cdot 10^4$.
  The optimal MSE jumps at
  $\alpha_{\rm opt}$. Hence for $\epsilon<0.00075$ there is a range of
  measurement rates $(\alpha_{\rm opt,\alpha_{\rm BP}})$ in which the
  BP is asymptotically suboptimal.}
\label{fig_MSE2}
\end{figure}

In Fig.~\ref{fig_MSE2} we plot in full blue line the MSE to which the
G-AMP converges and compare to the MSE achieved by the optimal Bayes
inference, i.e. the MSE corresponding to the global maximum of $\Phi(E)$ (in dashed red line). 
We see that, when the discontinuous transition point $\alpha_{\rm BP}$
exists, then in the region $\alpha_{\rm opt}<\alpha<\alpha_{\rm BP}$
G-AMP is suboptimal. We remind that, in the limit$\epsilon \to 0$, exact
reconstruction is possible for any $\alpha>\rho$. We see that for
$\alpha<\alpha_{\rm opt}$ and for $\alpha>\alpha_{\rm BP}$ the
performance of G-AMP matches asymptotically the performance of the
optimal Bayes inference. The two regions are, however, quite
different. For $\alpha<\alpha_{\rm opt}$  the final MSE is relatively
large, whereas for $\alpha>\alpha_{\rm BP}$ the final MSE is of order
$\epsilon$ and hence is this region the problem shows a very good
stability towards approximate sparsity. 

\begin{figure}[!ht]
\includegraphics[width=3.5in]{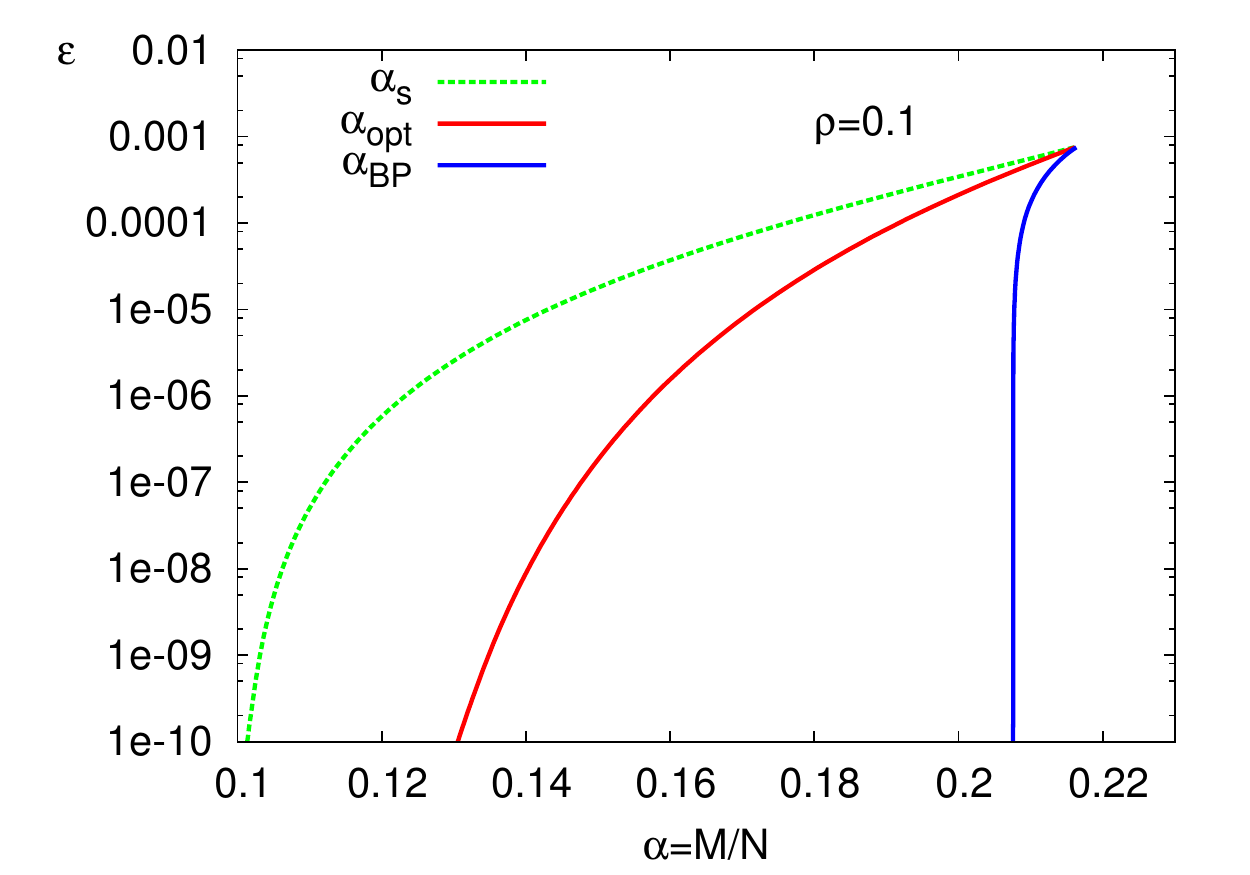}
\caption{Phase diagram for compressed sensing of approximately sparse
  signals. The density of the large signal components is $\rho=0.1$,
we are changing measurement rate $\alpha$ and variance of small components
  $\epsilon$.  The critical values of measurement rates $\alpha_{\rm opt}$, $\alpha_{BP}$
  and $\alpha_s$ are plotted. For homogeneous measurement matrices, BP
  does not achieve optimal reconstruction in the area between
  $\alpha_{\rm opt}$ (red) and $\alpha_{\rm BP}$ (blue). 
}
\label{fig_phase_diagram}
\end{figure}

In Fig.~\ref{fig_phase_diagram} we summarize the critical values of
$\alpha_{\rm BP}$ and $\alpha_{\rm opt}$ for a signal of density
$\rho=0.1$ as a function of the variance of the small components
$\epsilon$. Note that for $\epsilon>0.00075$ (the value depends on
$\rho$) there are no phase transitions, hence for this large value of
$\epsilon$, the G-AMP algorithm matches asymptotically the optimal
Bayes inference.  Note that in the limit of exactly sparse signal
$\epsilon\to 0$ the values $\alpha_{\rm opt}\to \rho$, and
$\alpha_{\rm s}\to \rho$. Whereas $\alpha_{\rm BP} (\epsilon \to 0)\to
0.2076$, hence for $\alpha>0.2076$ the G-AMP algorithm is very robust
with respect to appearance of approximate sparsity since the
transition $\alpha_{\rm BP}$ has a very weak $\epsilon$-dependence, as
seen in Fig.~\ref{fig_phase_diagram}.

\begin{figure*}[!ht]
\includegraphics[width=2.9in]{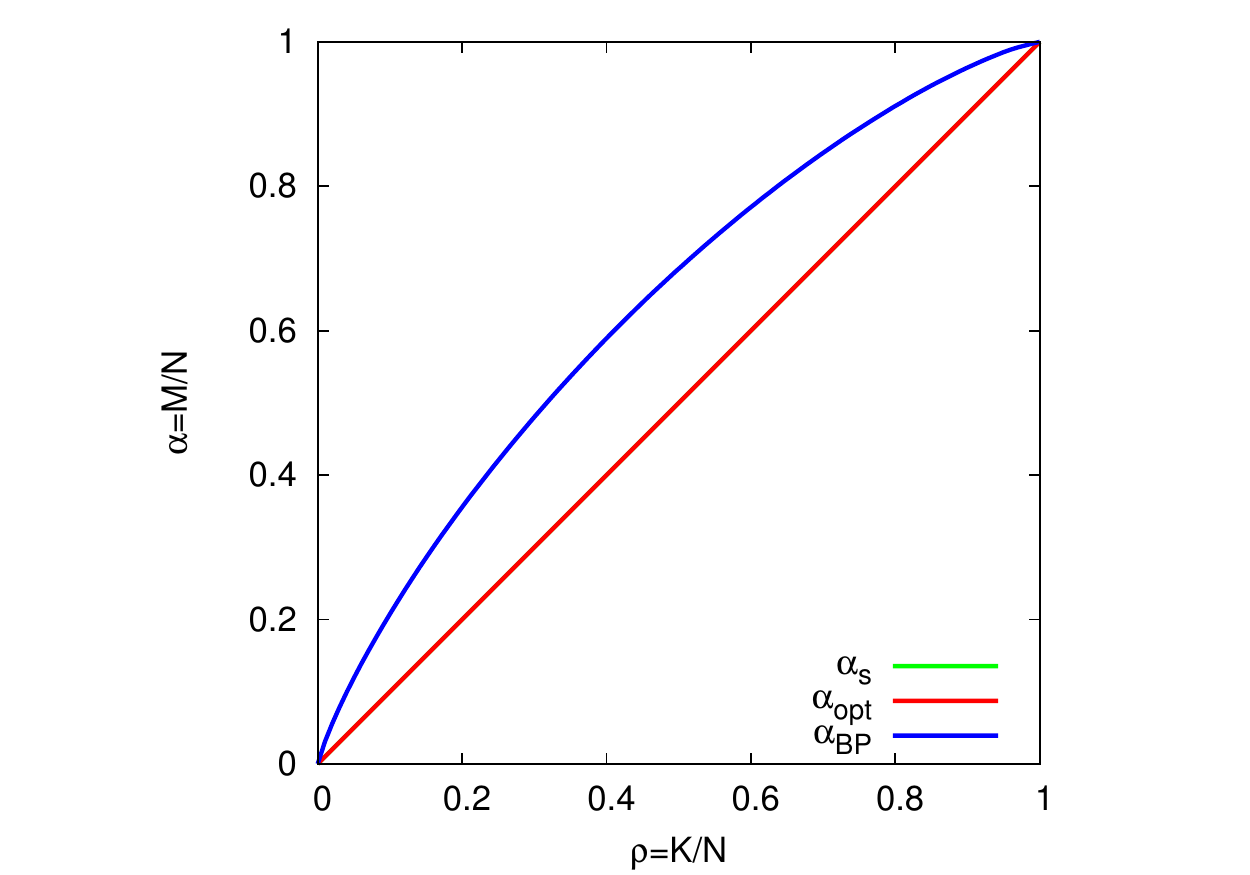}\hspace{-2cm}
\includegraphics[width=2.9in]{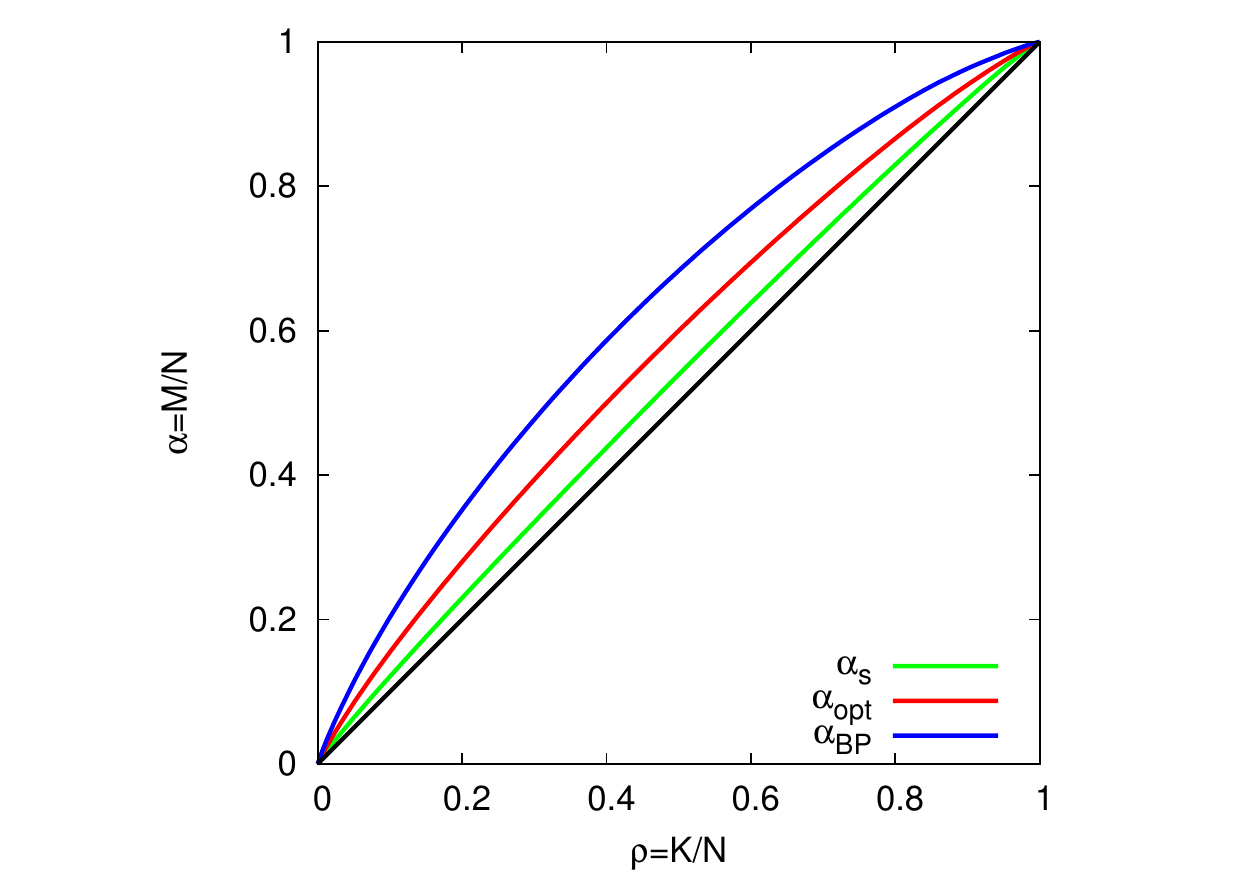}\hspace{-2cm}
\includegraphics[width=2.9in]{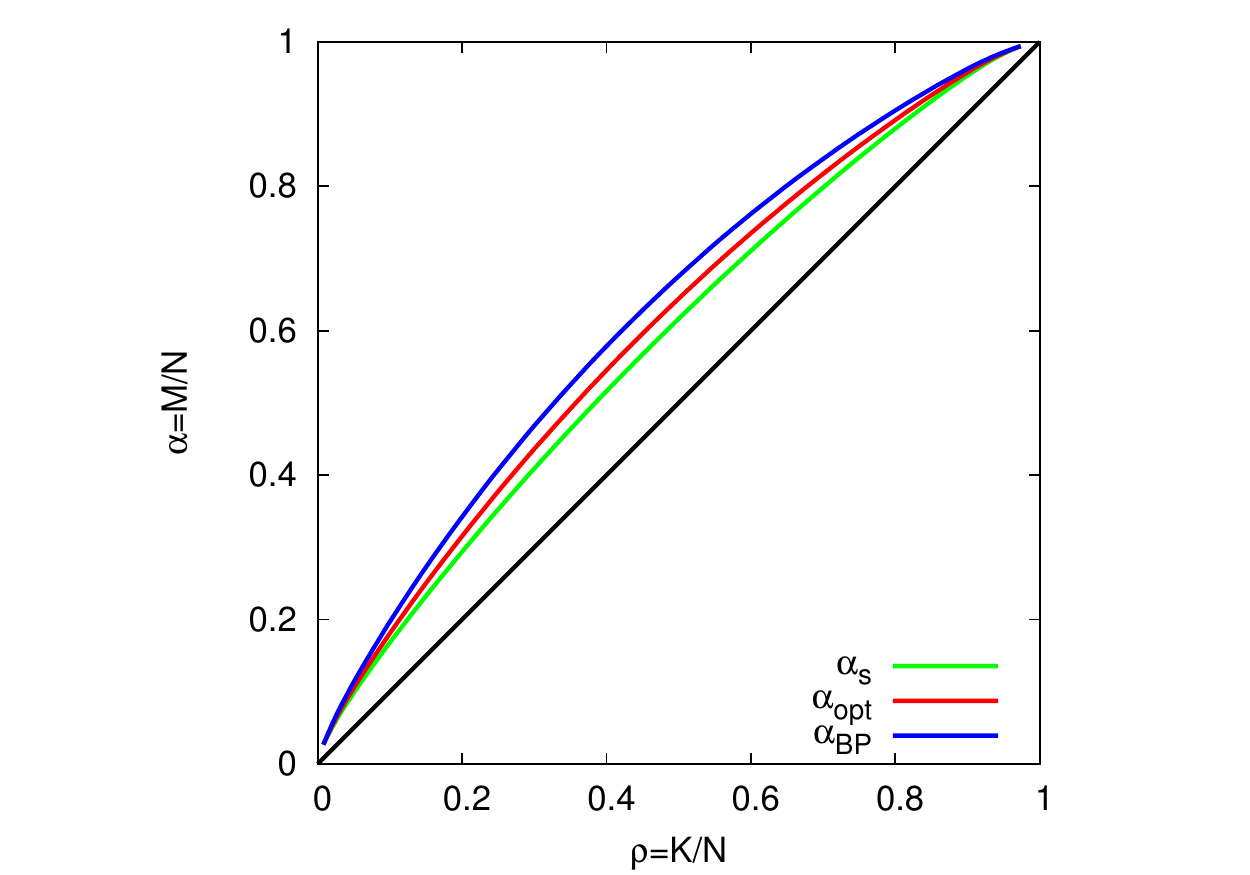}
\caption{Phase diagram in the plane density $\rho$ measurement rate
  $\alpha$,  with variance of small components $\epsilon=0$
  (left),  $\epsilon=10^{-6}$ (center) and  $\epsilon=10^{-4}$ (right).}
\label{fig_phase_general}
\end{figure*}

In Fig.~\ref{fig_phase_general} we plot the phase diagram for fixed
variance $\epsilon$ in the density $\rho$ - measurement rate $\alpha$
plane. The only
space for improvement is in the region $\alpha_{\rm opt}<\alpha
<\alpha_{\rm BP}$. In this region, G-AMP is not optimal because the
%M Small changes
potential $\Phi(E)$ has two maxima, and the 
 iterations
are blocked in the ``wrong''  local maximum of the potential $\Phi(E)$
with the largest $E$. This situation is well known in physics as a
first order transition, with a blocking of the dynamics in a
metastable state.

\section{Reconstruction of approximately sparse signals with optimality
achieving matrices}
\label{seeding}

A first order phase transition that is causing a failure
(sub-optimality) of the G-AMP algorithm appears also in the case of
truly sparse signals \cite{KrzakalaPRX2012}. In that case
\cite{KrzakalaPRX2012} showed that with the so-called ``seeding'' or
``spatially coupled'' measurement matrices the G-AMP algorithm is able
to restore asymptotically optimal performance. This was proven
rigorously in \cite{DonohoJavanmard11}. Using arguments from the
theory of crystal nucleation, it was argued heuristically in
\cite{KrzakalaPRX2012} that spatial coupling provides improvement
whenever, but only if, a first order phase transition is present. 
Spatial coupling was first suggested for compressed sensing in
\cite{KudekarPfister10} where the authors tested cases without a first
order phase transition (see Fig.~\ref{fig_MSE1}), hence no improvement was observed. Here we show
that for measurement rates $\alpha_{\rm opt}<\alpha <\alpha_{\rm BP}$
seeding matrices allow to restore optimality also for the inference of
approximately sparse signals.

\subsection{Seeding sampling matrices}

\begin{figure}[!ht]
\includegraphics[width=3.5in]{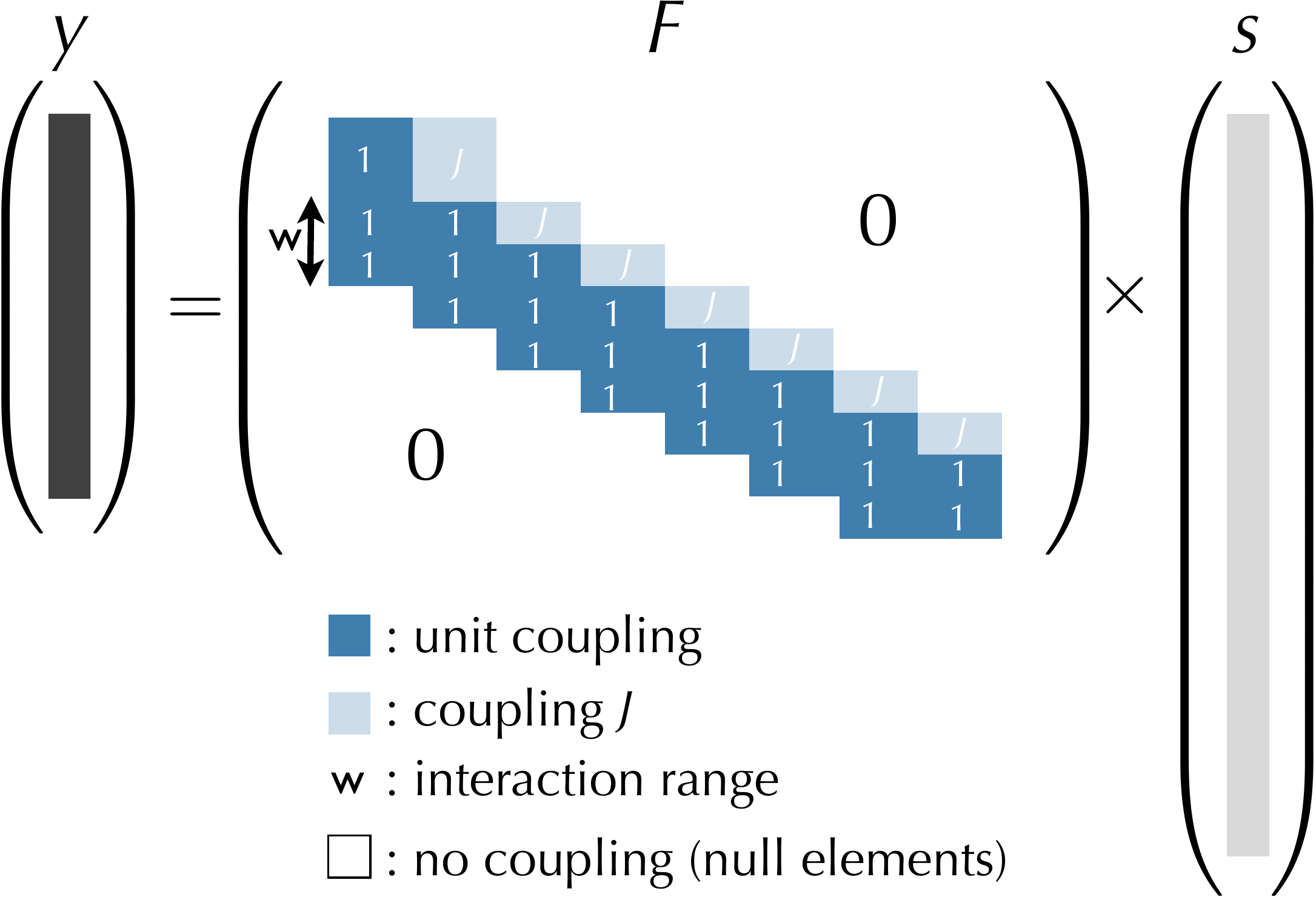}
\caption{Sketch of the measurement matrices we used to approach
  optimal reconstruction with the G-AMP algorithm. We use a number of
  variable-blocks $L_c$, and $L_r$ measurement blocks. The matrix
  components are iid with zero mean and variance $1/N$ for the blocks
  on the diagonal and for a number $W$ of lower diagonals, the upper
  diagonal blocks have components with variance $J/N$.}
\label{fig_matrix}
\end{figure}

The block measurement matrices $F_{\mu i}$ that we use in the rest of
this paper are constructed as follows: The $N$ variables are divided
into $L_c$ equally sized groups. And the $M$ measurements are divided
into $L_r$ groups of $M_{\rm seed}$ measurements in the first group
and $M_{\rm bulk}$ in the others.  We define $\alpha_{\rm seed}=L_c
M_{\rm seed}/N$ and $\alpha_{\rm bulk}=L_c M_{\rm bulk}/N$.  The total
measurement rate is \be \alpha= \frac{\alpha_{\rm seed} +(L_r-1)
  \alpha_{\rm bulk} }{L_c} \label{alpha} \ee The matrix $F$ is then
composed of $L_r\times L_c$ blocks and the matrix elements $F_{\mu i}$
are generated independently, in such a way that if $\mu$ is in group
$q$ and $i$ in group $p$ then $F_{\mu i}$ is a random number with zero
mean and variance $J_{q,p}/N$. Thus we obtain a $L_r\times L_c$
coupling matrix $J_{q,p}$. For the asymptotic analysis we assume that
$N\to \infty$ and $\alpha_{\rm seed}$, $\alpha_{\rm bulk}$ are
fixed. Note that not all block matrices are good seeding matrices, the
parameters have to be set in such a way that seeding is implemented
(i.e. existence of the seed and interactions such that the seed
grows). The coupling measurement matrix $J_{q,p}$ that we use in this
paper is illustrated in Fig.~\ref{fig_matrix}.

To study the state evolution for the block matrices we define
$E_p^{t}$ to be the mean-squared error in block $p$ at time $t$. The
$E_p^{t+1}$ then depends on $\hat m_p^t$ from the same block according to eq.~(\ref{Et}), on the
other hand the quantity $\hat m_p^t$ depends on the MSE $E^{t}_q$ from
all the blocks $q=1,\dots,L_c$ as follows
\be 
     \hat m_p^t  =  \frac{\alpha_{\rm seed}J_{1p}}{\sum_{q=1}^{L_c}
       J_{1q} E_q^t }  + \alpha_{\rm bulk}\sum_{r=2}^{L_r}
     \frac{J_{rp}}{\sum_{q=1}^{L_c} J_{rq} E_q^t } 
\ee

This kind of evolution belongs to the class for which threshold
saturation (asymptotic achievement of performance matching the optimal
Bayes inference solver) was proven in \cite{YedlaJian12} (when $L_c\to \infty$,
$W\to \infty$ and $L_c/W \gg 1$).
%M 
This asymptotic guarantee is reassuring, but one must check if finite
$N$ corrections are gentle enough to be able to
perform compressed sensing close to $\alpha_{\rm opt}$ even for
practical system sizes. We
hence devote the next section to numerical experiments showing that
the G-AMP
algorithm is indeed able to reconstruct close to optimality with
seeding matrices.  

\subsection{Restoring optimality}

\begin{figure}[!ht]
\includegraphics[width=3.5in]{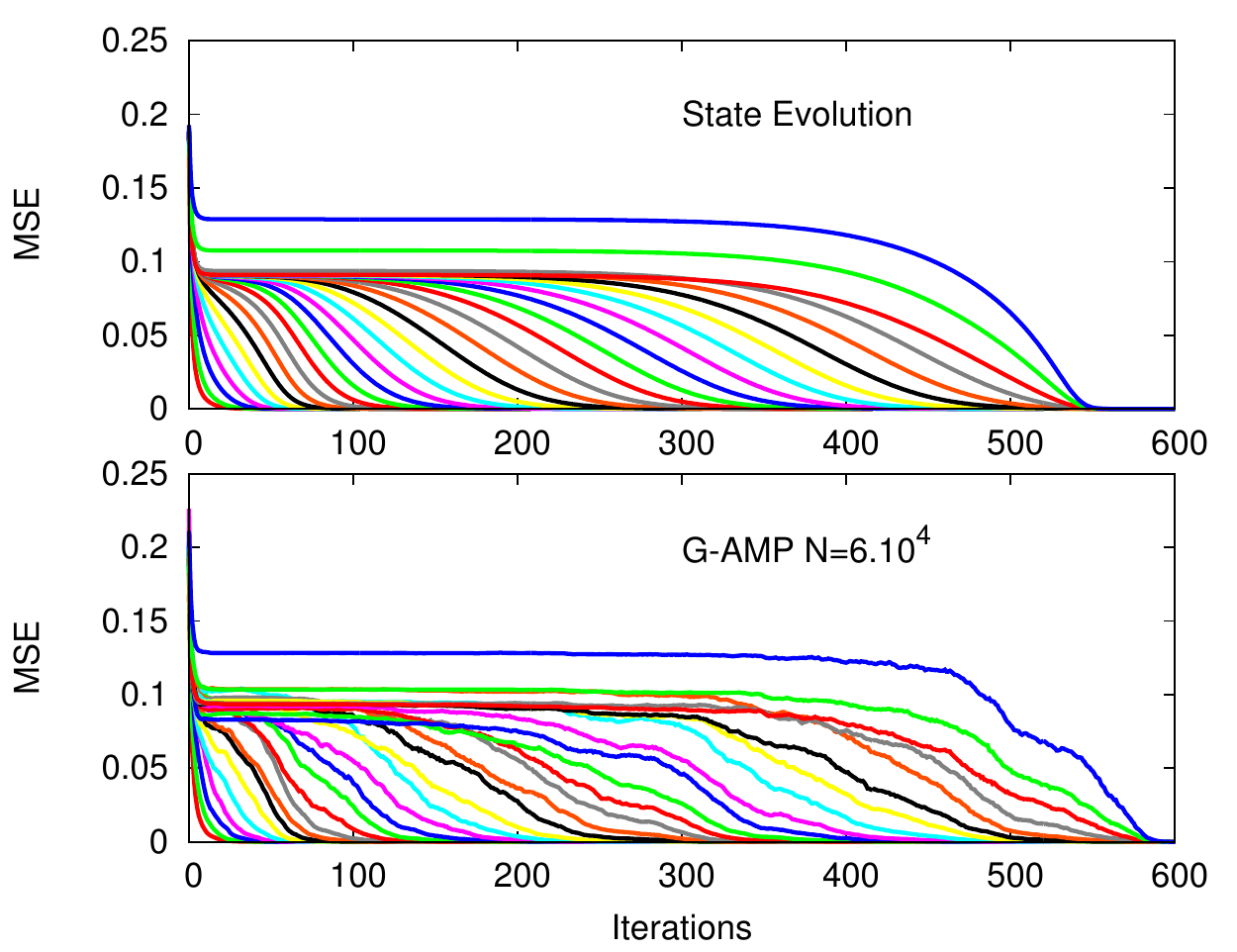}
\caption{Evolution of the MSE in reconstruction of signal with density
  $\rho=0.2$, variance of small components $\epsilon=10^{-6}$ at
  measurement rate $\alpha=0.303$. The state evolution on the top is
  compared to the evolution of the algorithm for a signal size
  $N=6\cdot 10^4$ on the bottom. The measurement is performed using a
  seeding matrix with the following parameters: $\alpha_{\rm seed}=0.4$, 
  $\alpha_{\rm bulk}=0.29$, $W=3$, $J=0.2$, $L_c =30$. 
}
\label{fig_wave}
\end{figure}

In Fig.~\ref{fig_wave} we show the state evolution compared to the
evolution of the G-AMP algorithm for system size
$N=6\cdot 10^4$. The signal was of density $\rho=0.2$ and
$\epsilon=10^{-6}$, the parameters of the measurement matrix are in
the second line of Table~\ref{param}, the $L_c=30$ giving measurement rate
$\alpha=0.303$ which is deep in the region where G-AMP for homogeneous
measurement matrices gives large MSE (for $\alpha<0.356$). We see finite size fluctuations, but overall the
evolution corresponds well to the asymptotic curve. 

\begin{figure}[!ht]
\includegraphics[width=3.in]{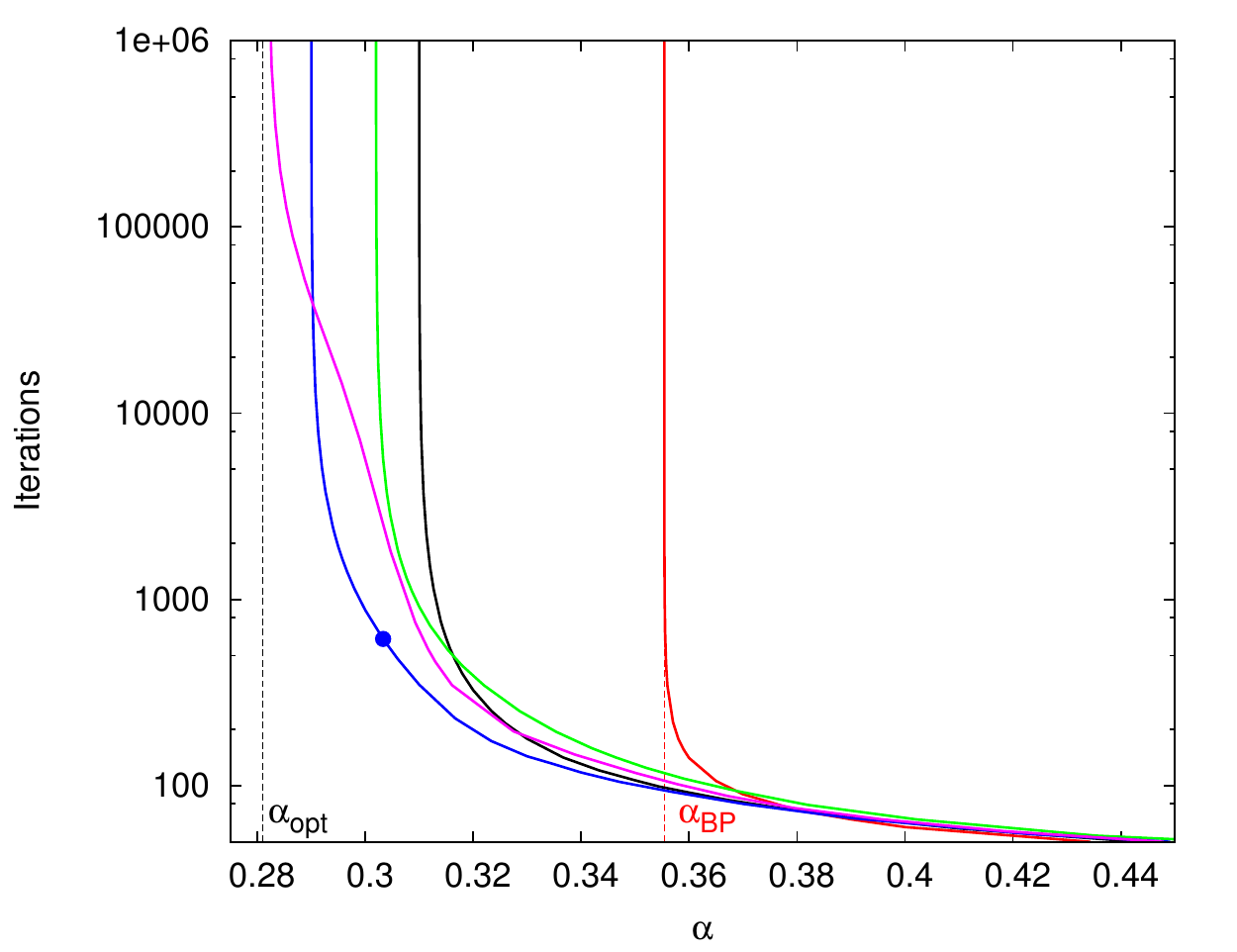}
\caption{The convergence time of BP for large system sizes estimated
  by the state evolution as a function of measurement rate
  $\alpha$. Data are for signals with density $\rho=0.2$, variance of
  small components $\epsilon=10^{-6}$.  The red line is obtained using
  an homogeneous measurement matrix, the vertical dashed line
  corresponds to the limit this approach can achieve $\alpha_{\rm
    BP}=0.3554$. All the other lines are obtained using seeding matrix
  with parameters specified in Table~\ref{param} and varying $L_c$,
  the resulting measurement rate $\alpha$ is computed from
  eq. (\ref{alpha}).  With these seeding matrices and using large
  $L_c$, reconstruction is possible at least down to $\alpha_{\rm
    bulk}=0.282$ which is very close to the measurement rate
  $\alpha_{\rm opt}=0.2817$. The blue point corresponds to the
  evolution illustrated in Fig.~\ref{fig_wave}.}
\label{fig_time}
\end{figure}

\begin{table}[!ht]
\begin{center}
\begin{tabular}{|c|c|c|c|c|c|c||}
  \hline
  color & $\alpha_{\rm seed}$ & $\alpha_{\rm bulk}$ & $J$ & $W$ & $L_r$ \\
  \hline
  violet & $0.4$ & $0.282$ & $0.3$ & $3$ & $L_c+2$ \\
   blue & $0.4$ & $0.290$ & $0.2$ & $3$ & $L_c+1$ \\
   green & $0.4$ & $0.302$ & $0.001$ & $2$ & $L_c+1$ \\
   black & $0.4$ & $0.310$ & $0.4$ & $3$ & $L_c+1$\\
\hline
\end{tabular}
\end{center}
\caption{Parameters of the seeding matrices used in Fig.~\ref{fig_time}
  \label{param}}
\end{table}

In Fig.~\ref{fig_time} we plot the convergence time needed to achieve
reconstruction with $E \approx \epsilon$ for several sets of
parameters of the seeding matrices. With a proper choice of the
parameters, we see that we can reach an optimal reconstruction for
values of $\alpha$ extremely close to $\alpha_{\rm opt}$. Note, however,
that the number of iterations needed to converge diverges as $\alpha
\to \alpha_{\rm opt}$. This is very similar to what has been obtain in the
case of purely sparse signals in \cite{KrzakalaPRX2012,DonohoJavanmard11}.

\begin{figure}[!ht]
\includegraphics[width=3.in]{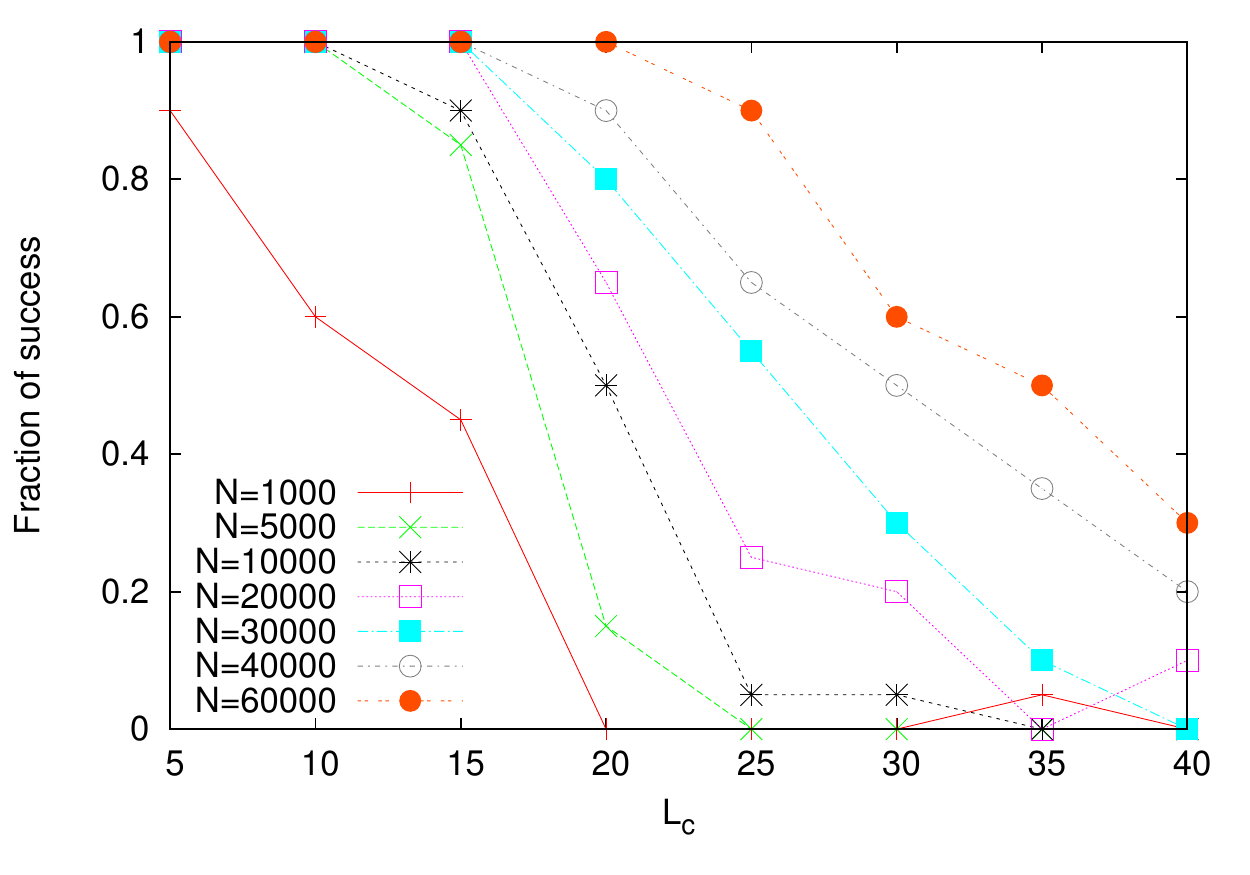}
\caption{Fraction of instances (over $20$ attempts) that were solved
  by the algorithm in less than twice the number of iterations
  predicted by the density evolution for different system sizes, as a
  function of the number of blocks $L_c$. We used the parameters that
  lead to the blue curve in Fig.~\ref{fig_time} (i.e. second line of
  Table~\ref{param}). As $N \to \infty$, reconstruction is reached in
  all the instances, as predicted by the state evolution. For finite
  $N$, however, reconstruction is not reached when $L_c$ is too
  large.}
\label{convergence}
\end{figure}

Finally, it is important to point out that this theoretical analysis
is valid for $N \to \infty$ only. Since we eventually work with finite
size signals, in practice, finite size effects slightly degrade this
asymptotic threshold saturation. 
This is a well known effect in coding theory where a major question is
how to optimise finite-length codes (see for instance
\cite{AmraouiMontanari2007}).
In Fig.~\ref{convergence} we plot the fraction of cases in which the
algorithm reached successful reconstruction for different system sizes
as a function of the number of blocks $L_c$. We see that for a given
size as the number of blocks is growing, i.e. as the size of one block
decreases, the performance deteriorates. As expected the situation
improves when size augments. Analyses of the data presented in
Fig.~\ref{convergence} suggest that the size of one block that is
needed for good performance grows roughly linearly with the number of
blocks $L_c$.
% We saw $N_{\rm block} = 80 L - 1000$. 
This suggests that the probability of failure to transmit the
information to every new block is roughly inversely proportional to
the block size. We let for future work a more detailed investigation
of these finite size effects. The algorithm nevertheless reconstructs
signals at rates close to the optimal one even for system sizes of
practical interest.

\section{Discussion}

\subsection{Mismatching signal model} 

In this paper we treated signals generated by the two-Gaussian model
and we assumed knowledge of the parameters used for the
generation. Note, however, that in the same way as in
\cite{KrzakalaPRX2012,VilaSchniter11,KrzakalaMezard12} the
corresponding parameters ($\rho$, $\epsilon$, etc.) can be learned using expectation
maximization. For real data it is also desirable to use a signal model
(\ref{Px}) that is matching the data in a better way. 

At this point we want to state that whereas all our results do
depend quantitatively on the statistical properties of the signal,
the qualitative features of our results (e.g. the presence and nature of
the phase transitions) are valid for other signals,
distinct from the two-Gaussian case that we have studied here, and even for the case when the
signal model does not match the statistical properties of the actual
signal.  This was illustrated e.g. for the noisy compressed sensing of
truly sparse signal in \cite{KrzakalaMezard12}. In the same line,
we noticed and tested that if G-AMP corresponding to $\epsilon=0$ is run for the approximately
sparse signals the final MSE is always larger than the one achieved by
G-AMP with the right value of $\epsilon$. 

We tested the G-AMP algorithm with the signal model (\ref{Px}) and
EM learning of  parameters on some real images and we indeed observed better performance than
for the G-AMP designed for truly sparse signals. However, to become
competitive we also need to find better models for the signal, likely
including the fact that the sparse components are highly structured
for real images. We let this for future work. 

\subsection{Presence of noise} 

In this paper we studied the case of noiseless measurements, but the
measurement noise can be straightforwardly included into the analysis as in
\cite{KrzakalaMezard12} where we studied the phase diagram in the
presence of the measurement noise. The results would change
quantitatively, but not qualitatively.

\subsection{Computational complexity}
The G-AMP algorithm as studied here runs in $O(MN)$ steps. For dense
random matrices this cannot be improved, since we need $MN$ steps to
only read the components of the matrix. Improvements are possible for
matrices that permit fast matrix operations, e.g. Fourrier or Gabor
transform matrices \cite{JavanmardMontanari12}.
 Again, testing
approximate sparsity in this case is an interesting direction for
future research. Note, however, that the state evolution and replica
analysis of optimality does not apply (at least not straightforwardly)
to this case.

Another direction to improve the running time of G-AMP is to sample
the signal with sparse matrices as e.g. in
\cite{BaronSarvotham10}. For sparse matrices G-AMP is not anymore
asymptotically equivalent to the belief propagation (even though it
can be a good approximation), and the full belief propagation
is much harder to implement and to analyze. But despite this
difficulty, it is an interesting direction to investigate.

\subsection{Spatial coupling}

For small variance of the small components of the signal the G-AMP
algorithm for homogeneous matrices does not reach optimal
reconstruction for measurement rates close to the theoretical limit
$\alpha_{\rm opt}$. The spatial coupling approach, resulting in the
design of seeding matrices, improves significantly the
performance. For diverging system sizes optimality can be restored. We
show that significant improvement is also reached for sizes of
practical interest. There are, however, significant finite size effect
that should be studied in more detail. The optimal design of the
seeding matrix for finite system sizes (as studied for instance in
depth in the context of error correcting codes
\cite{AmraouiMontanari2007}) remains an important open
question.

%\section{Acknowledgements}

\bibliographystyle{IEEEtran}
\bibliography{IEEEabrv,refs}

\end{document}